\documentclass[11pt,a4paper]{article}
\usepackage{amsmath}
\usepackage[mathcal]{eucal}
\usepackage{latexsym}
\usepackage{amssymb}
\begin{titlepage}
\title{Finite states in four dimensional quantum gravity. The isotropic minisuperspace Asktekar--Klein--Gordon model.}
\author{Eyo Eyo Ita III}
\input amssym.def
\input amssym.tex

\begin{document}
\maketitle
\bigskip
\centerline{Department of Applied Mathematics and Theoretical Physics} 
\smallskip
\centerline{Centre for Mathematical Sciences, University of Cambridge, Wilberforce Road}
\smallskip
\centerline{Cambridge CB3 0WA, United Kingdom}
\smallskip
\centerline{eei20@cam.ac.uk} 

\bigskip

\begin{abstract}
\par
\medskip
\indent
In this paper we construct the generalized Kodama state for the case of a Klein--Gordon scalar field coupled to Ashtekar variables in isotropic minisuperspace by a new method.  The criterion for finiteness of the state stems from a minisuperspace reduction of the quantized full theory, rather than the conventional techniques of reduction prior to quantization.  We then provide a possible route to the reproduction of a semiclassical limit via these states.  This is the result of a new principle of the semiclassical-quantum correspondence (SQC), introduced in the first paper in this series.  Lastly, we examine the solution to the minisuperspace case at the semiclassical level for an isotropic CDJ matrix neglecting any quantum corrections and examine some of the implications in relation to results from previous authors on semiclassical orbits of spacetime, including inflation.  It is suggested that the application of nonperturbative quantum gravity, by way of the SQC, might potentially lead to some predictions testable below the Planck scale.
\end{abstract}
\end{titlepage}

\section{Introduction: Reduction to minisuperspace}

\noindent
In \cite{EYO} we provided a new method for constructing finite states of quantum gravity.  The underlying premise is that in order to canonically quantize a general theory in congruity with the axioms of quantum field theory, as a bare minimum all of the canonical commutation relations defining the field theory must be exhaustively and consistently applied.  There are some additional conditions which we will not go into in this work but ultimately, reproduction of a good semiclassical limit might serve as a possible check on the viability of any method.  Our main focus here, as in \cite{EYO}, is the exhaustive application of the canonical commutation relations as well as to present some possible routes toward observational verification of the semiclassical limit.\par
\indent  
The commutation relations of any quantum field theory must be defined, in the $3+1$ description, with respect to each spatial hypersurface $\Sigma_t$ labeled by time $t$ for $t_0\leq{t}\leq{T}$.  In the case of gravity, the arena for quantization is the configuration space of Ashtekar connections and matter fields $\Gamma=(A^a_i,\phi^{\alpha})$ living on a three-dimensional hypersurface $\Sigma_t$ within a four-dimensional 
manifold $M=\Sigma\otimes{R}$.  The nontrivial commutation relations $[\widetilde{\sigma}^a_i,A^b_j]$, $[\pi,\phi]$ applied at the same point induce quantum singularities which by the 
semiclassical-quantum correspondence were made in \cite{EYO} to vanish, producing a unique quantum state out of a two-parameter family of WKB solutions.  This is a nontrivial condition which contains physical content.  The trivial commutation relations $[\widetilde{\sigma}^a_i,\pi]$ established the mixed partials condition, also with nontrivial physical content as it provides a functional boundary condition on the semiclassical matter mometum labeling the generalized Kodama state, which in this paper we will denote by the function $f=f(\phi)$.  The remaining trivial relations $[\widetilde{\sigma}^a_i,\phi]$ and $[\pi,A^a_i]$ are also rich in physical content since they allowed for the explicit construction of the wavefunction by way of a new functional calculus, exploiting the dynamical independence of the variables.\par
\indent
These results were presented in \cite{EYO} for the full theory.  A fundamental question in the quantization of gravity is whether the process of quantization and minisuperspace reduction commute.  We claim that one must reduce the fully quantized theory in order to obtain correct results rather than the opposite since, in the former case full use is made by our method of all the aformentioned canonical commutation relations with their associated physical content.  It suffices to reduce the 
equations $q_0=q_1=q_2=0$ derived in \cite{EYO} in order to meet these requirements for minisuperspace.\par
\indent
We illustrate in this paper as a simple example the solution of the constraints under symmetry reduction and the construction of the corresponding generalized Kodama state in the presence of a homogeneous Klein--Gordon scalar field in isotropic minisuperspace.  The first four sections are devoted to reducing the conditions derived in \cite{EYO} from quantization of the full theory into minisuperspace, starting with the kinematic constraints and then considering the mixed partials condition and the Hamiltonian constraint.  In the minisuperspace case the kinematic constraints become trivialized leaving just the Hamiltonian constraint, which we explicitly solve for the ingredients necessary to construct the generalized Kodama state.  We construct the state and provide a physical interpretation.  The purpose is to demonstrate, starting with the simplest nontrivial example, some features of relevance to the full theory as well as to indicate a possible route toward observational verification of the semiclassical limit prior to moving on to more complicated examples.\par
\indent  
In the last section we treat the solution of the Hamiltonian constraint at just the semiclassical level and analyse the semiclassical orbits of the corresponding spacetime.  We obtain some intuition on the effects of inflation due to the presence of the Klein--Gordon field.  The net result is a field-dependent cosmological constant which plays the role of the CDJ matrix.  The entering assumption in this case and in the full theory, as for all generalized Kodama states, is nondegeneracy of the Ashtekar magnetic field.\par
\indent
For this work, all definitions and conventions are used in accordance with \cite{EYO} as applied to minisuperspace.

\section{Kinematic constraints in minisuperspace}
\par
\medskip

The CDJ matrix, though spatially homogeneous in minisuperspace, must still satisfy the kinematic constraints of general relativity in the Ashtekar variables as a necessary condition for completeness of the quantization procedure.  Starting with Gauss' law \cite{EYO},

\begin{equation}
\label{MINI}
B^i_{e}D_{i}\Psi_{ae}=\partial_{i}\Psi_{ae}+B^i_{e}A^b_{i}\bigl(f_{abc}\Psi_{ce}
+f_{ebc}\Psi_{ae}\bigr)=GQ_{a}.
\end{equation}

\noindent
We will make repeated use of the following identities regarding the Ashtekar magnetic field

\begin{eqnarray}
\label{MINI1}
B^i_{e}=\epsilon^{ijk}F^e_{jk}=\epsilon^{ijk}\bigl(\partial_{j}A^e_{k}
+{1 \over 2}f^{edf}A^d_{j}A^f_{k}\bigr)
=\epsilon^{ijk}\partial_{j}A^e_{k}+{1 \over 2}\epsilon^{ijk}\epsilon_{edf}A^d_{j}A^f_{k}\nonumber\\
=\epsilon^{ijk}\partial_{j}A^e_{k}+(A^{-1})^i_{e}\hbox{det}A=(A^{-1})^i_{e}\hbox{det}A~in~minisuperspace.
\end{eqnarray}

\noindent
Making use of (\ref{MINI1}) and the identity

\begin{eqnarray}
\label{MINI3}
B^i_{e}A^b_{i}=A^b_{i}\epsilon^{ijk}\partial_{j}A^e_{k}+\hbox{det}A(A^{-1})^i_{e}A^b_{i}
=\delta^b_{e}\hbox{det}A~in~minisuperspace.
\end{eqnarray}

\noindent
One has, simplifying Gauss' law in (\ref{MINI})

\begin{eqnarray}
\label{MINI4}
B^i_{e}D_{i}\Psi_{ae}=\hbox{det}A\bigl[f_{abc}(A^{-1})^i_{e}A^b_{i}\Psi_{ce}
+f_{ebc}(A^{-1})^i_{e}A^b_{i}\Psi_{ae}\bigr]\nonumber\\
=\hbox{det}A\bigl[f_{abc}\delta^b_{e}\Psi_{ce}+f_{ebc}\delta^b_{e}\Psi_{ac}
=\hbox{det}A\bigl[f_{aec}\Psi_{ce}+f_{eec}\Psi_{ae}\bigr].
\end{eqnarray}

\noindent
The last term on the last line of (\ref{MINI4}) vanishes due to antisymmetry of the structure constants $f_{abc}$, and we are left with

\begin{equation}
\label{MINI5}
B^i_{e}D_{i}\Psi_{ae}=f_{aec}\Psi_{ce}=GQ_{a}.
\end{equation}

\noindent
Equation (\ref{MINI5}) is a statement that the antisymmetric part of the CDJ matrix in minisuperspace is determined by the $SU(2)_{-}$ charge.\par
\indent  
There are two implications
concerning this: (i) Only for a matter field transforming nontrivially under $SU(2)_{-}$ will 
(\ref{MINI5}) yield a nontrivial right hand side.  This includes fermions and excludes Klein--Gordon scalars. (ii) Recall that the antisymmetric part of the CDJ matrix also is in general determined by the matter contribution to the diffeomorphism constraint \cite{THIE2}

\begin{eqnarray}
\label{MINI6}
\epsilon_{dae}\Psi_{dae}=G{{B^i_{d}H_i} \over {\hbox{det}B}}
=G{{(A^{-1})^i_{d}H_i} \over {\hbox{det}A}}~in~minisuperspace.
\end{eqnarray}

\noindent
This implies a relationship between the matter contribution to the diffeomorphism constraint and the Gauss' law constraint, given by

\begin{eqnarray}
\label{MINI7}
H_{i}=A^a_{i}Q_{a}\longrightarrow{Q}_a=(A^{-1})^i_{a}H_i
\end{eqnarray}

\noindent
Equation (\ref{MINI7}) highlights a kind of duality between SU(2) gauge transformations and diffeomorphisms.  Note that the connection $A^a_i$ must be nondegenerate in order for this to be the case.  This is equivalent to the requirement that the magnetic field $B^i_a$ be nondegenerate, and is so due to the presence of the matter field and the one-to-one map from 
$A^a_i$ to $B^i_a$ (\ref{MINI3}).  The corresponding relationship in the full theory is given by

\begin{eqnarray}
\label{MINI8}
H_{i}=\pi_{\alpha}\partial_{i}\phi^{\alpha}+A^a_{i}Q_{a}
\end{eqnarray}

\subsection{The Klein--Gordon scalar field in minisuperspace}

\indent
In the case of the Klein--Gordon scalar field, we have $Q_a=0$.  Due to the relation 
(\ref{MINI7}), this implies that $H_i=0$.  In other words, for the Klein--Gordon model, the
Gauss' law and diffeomorphism constraints are not independent.  Since there are nine CDJ matrix elements $\Psi_{ae}$, the number of degrees of freedom reduces from nine to six upon application of the six kinematic constraints.  This leaves six degrees of freedom for the quantized Hamiltonian constraint $(q_0=q_1=q_2=0)$ which upon solution leaves three CDJ matrix elements freely specifiable.  Hence, for the minisuperspace Klein--Gordon scalar model, three degrees of freedom are freely specifiable after solving all of the quantum constraints.  This is in contrast with the full theory.  In this case the diffeomorphisms are independent of the gauge transformations due to the spatial gradient term in (\ref{MINI8}).\par
\indent
For $Q_a=0$, (\ref{MINI5}) reads

\begin{eqnarray}
\label{MINI9}
B^i_{e}D_{i}\Psi_{ae}=f_{aec}\Psi_{ce}=0.
\end{eqnarray}

\noindent
This is a statement that the antisymmetric part of the CDJ matrix is zero, or that the CDJ matrix is symmetric.  Note that this is consistent with minisuperspace, as the scalar
field contribution to the diffeomorphism constraint also vanishes.  The CDJ matrix can then be parametrized by 

\begin{eqnarray}
\label{MINI10}
\Psi_{ab}=O_{ac}(\vec\theta)D_{cd}O^{T}_{db}(\vec\theta)
\end{eqnarray}

\noindent
where in (\ref{MINI10}), $\vec\theta$ is a set of three $SU(2)_{-}$ rotation angles and $D_{cd}$ is a diagonal matrix derivable from the CDJ matrix by orthogonal transformation. 
We shall pick $\vec\theta=0$ for simplicity.  This corresponds to eliminating the shear components of the CDJ matrix, leaving just the anisotropy components to be solved for. 
This is in a sense the same as fixing a gauge.  Note that in the full theory there is no such freedom in the solution, which highlights an important difference from 
minisuperspace.  So we shall set

\begin{eqnarray}
\label{MINI11}
\Psi_{23}=\Psi_{32}=0;~~\Psi_{13}=\Psi_{31}=0;~~\Psi_{12}=\Psi_{21}=0.
\end{eqnarray}

\section{Mixed partials consistency condition}

A consistency condition on the solution of the constraints is the mixed partials condition which requires trivial canonical commutations between the matter and gravitation quantized conjugate momenta, given by

\begin{eqnarray}
\label{MINI12}
\hbar{G}{{\partial\pi} \over {\partial{A}^a_i}}
=-i\hbar{B}^i_{e}{{\partial\Psi_{ae}} \over {\partial\phi}}.
\end{eqnarray}

\noindent
Let us examine the restrictions upon $A^a_i$ imposed by a diagonal CDJ matrix.  Writing out the individual components

\begin{eqnarray}
\label{MINI14}
{{\partial\pi} \over {\partial{A^1_i}}}
=-{i \over G}B^i_{1}{{\partial\Psi_{11}} \over {\partial\phi}};~~
{{\partial\pi} \over {\partial{A^2_i}}}
=-{i \over G}B^i_{2}{{\partial\Psi_{22}} \over {\partial\phi}};~~
{{\partial\pi} \over {\partial{A^3_i}}}=
-{i \over G}B^i_{3}{{\partial\Psi_{33}} \over {\partial\phi}}
\end{eqnarray}

\noindent
In minisuperspace (\ref{MINI4}) translates into

\begin{eqnarray}
\label{MINI15}
{{\partial\pi} \over {\partial{A^1_i}}}
=-{i \over G}\vert{A}\vert(A^{-1})^i_{1}{{\partial\Psi_{11}} \over {\partial\phi}};\nonumber\\
{{\partial\pi} \over {\partial{A^2_i}}}
=-{i \over G}\vert{A}\vert(A^{-1})^i_{2}{{\partial\Psi_{22}} \over {\partial\phi}};\nonumber\\
{{\partial\pi} \over {\partial{A^3_i}}}
=-{i \over G}\vert{A}\vert(A^{-1})^i_{3}{{\partial\Psi_{33}} \over {\partial\phi}}
\end{eqnarray}

\noindent
Making use of the nondegeneracy of $A^a_i$, contract (\ref{MINI15}) by $A^k_i$ to yield the following condition

\begin{eqnarray}
\label{MINI16}
A^k_{i}{{\partial\pi} \over {\partial{A}^1_i}}
=-{i \over G}\vert{A}\vert\delta^k_{1}{{\partial\Psi_{11}} \over {\partial\phi}};\nonumber\\
A^k_{i}{{\partial\pi} \over {\partial{A}^2_i}}
=-{i \over G}\vert{A}\vert\delta^k_{2}{{\partial\Psi_{22}} \over {\partial\phi}};\nonumber\\
A^k_{i}{{\partial\pi} \over {\partial{A}^3_i}}
=-{i \over G}\vert{A}\vert\delta^k_{3}{{\partial\Psi_{33}} \over {\partial\phi}}
\end{eqnarray}

\noindent
Writing out (\ref{MINI16}) in individual components we have, for $k=1$,

\begin{eqnarray}
\label{MINI17}
A^1_{i}{{\partial\pi} \over {\partial{A}^1_i}}=
-{i \over G}\vert{A}\vert\delta^k_{1}{{\partial\Psi_{11}} \over {\partial\phi}};\nonumber\\
A^1_{i}{{\partial\pi} \over {\partial{A}^2_i}}=0;\nonumber\\
A^1_{i}{{\partial\pi} \over {\partial{A}^3_i}}=0
\end{eqnarray}

\noindent
Looking for example at the middle term in (\ref{MINI17}),  One has two possibilities:

\begin{eqnarray}
\label{MINI18}
A^1_{2}=0;~~{{\partial\pi} \over {\partial{A}^2_2}}~arbitrary,~or\nonumber\\
A^1_{2}~arbitrary;~~{{\partial\pi} \over {\partial{A}^2_2}}=0\nonumber\\
\end{eqnarray}

\noindent
Applying the analogous criterion to the counterparts to (\ref{MINI16})

\begin{eqnarray}
\label{MINI19}
A^2_{i}{{\partial\pi} \over {\partial{A}^1_i}}=0;~~
A^2_{i}{{\partial\pi} \over {\partial{A}^2_i}}
=-{i \over G}\vert{A}\vert{{\partial\Psi_{22}} \over {\partial\phi}};~~
A^2_{i}{{\partial\pi} \over {\partial{A}^3_i}}=0\nonumber\\
A^3_{i}{{\partial\pi} \over {\partial{A}^1_i}}=0;~~
A^3_{i}{{\partial\pi} \over {\partial{A}^2_i}}=0;~~
A^3_{i}{{\partial\pi} \over {\partial{A}^3_i}}
=-{i \over G}\vert{A}\vert{{\partial\Psi_{33}} \over {\partial\phi}}
\end{eqnarray}

\noindent
Equations (\ref{MINI17}) and (\ref{MINI19}) state that the matter conjugate momentum $\pi$ is independent of six directions in connection space.  Therefore it depends upon three independent directions $A^1_1=a_1$, $A^2_2=a_2$, $A^3_3=a_3$.  A diagonal CDJ matrix $\Psi_{ae}$ in minisuperspace implies a diagonal connection $A^a_i$.  Using $\vert{A}\vert=a_{1}a_{2}a_{3}$ and
making the definition $a_{i}=\hbox{exp}(\xi_i)$ for $i=1,2,3$, one can write the mixed partials condition as 

\begin{eqnarray}
\label{MINI20}
{{\partial\pi} \over {\partial\xi_1}}=-{i \over G}\hbox{exp}[\xi_1+\xi_2
+\xi_3]{{\partial\Psi_{11}} \over {\partial\phi}};\nonumber\\
{{\partial\pi} \over {\partial\xi_2}}=-{i \over G}\hbox{exp}[\xi_1+\xi_2
+\xi_3]{{\partial\Psi_{22}} \over {\partial\phi}};~~
{{\partial\pi} \over {\partial\xi_3}}=-{i \over G}\hbox{exp}[\xi_1+\xi_2
+\xi_3]{{\partial\Psi_{33}} \over {\partial\phi}}
\end{eqnarray}

\noindent
Note that if $a_1=a_2=a_3$, then it follows that $\Psi_{11}=\Psi_{22}=\Psi_{33}$ to within some arbitrary functions $f_i=f_i(a)$ of $a$ for $i=1,2,3$.  Hence, an isotropic CDJ matrix fixes the connection to be isotropic, and vice versa.  Note that this condition is unique to minisuperspace and not the full theory.\par

\section{Reduction of the quantum constraints}
\indent
Now that we have dealt with the diffeomorphism and Gauss' law constraints by fixing a gauge in which the CDJ matrix is diagonal as well as the connection $A^a_i$, we will attempt to solve the Hamiltonian constraint and construct the corresponding generalized Kodama state 
$\Psi_{GKod}[A,\phi]$.  In this work we are performing minisuperspace reduction after quantization.  In the minisuperspace approximation we have $T_{ij}=0$.  The kinematic contribution to the quantized constraints can be eliminated and the remaining constraints reduce to the system

\begin{eqnarray}
\label{SYSGKODONE}
\hbox{det}B\bigl(GV\hbox{det}\Psi+Var\Psi\bigr)
+{{\pi^2} \over 2}=0;\nonumber\\
\epsilon_{ijk}\epsilon^{abc}D^{kj}_{cb}\Psi_{ae}B^i_{e}+
\epsilon_{ijk}\epsilon^{abc}{\partial \over {\partial{A}^a_i}}
\Bigl[B^k_{c}B^j_{e}\Psi_{be}+{{GV} \over 4}B^k_{e}B^j_{f}\Psi_{ce}\Psi_{bf}\Bigr]
-{i \over {2G}}{{\partial\pi} \over \partial\phi}=0;\nonumber\\
{{GV} \over 6}{\partial \over {\partial{A^a_i}}}{\partial \over {\partial{A^b_j}}}
(\epsilon_{ijk}\epsilon^{abc}B^k_{e}\Psi_{ce})+36=0
\end{eqnarray}

\noindent 
subject to the condition that the CDJ matrix $\Psi_{ae}$ is diagonal.  In (\ref{SYSGKODONE}) we have appended a factor of $G$ to the self-interaction potential $V$.  Since the mass dimensions are $[G]=-2$ and $[V]=4$, the quantity $GV$ is of the same mass dimension of the cosmological constant $[\Lambda=2]$, the contribution of which can be deemed already included (as in $V\rightarrow{V}+{\Lambda \over G}$).  We will further simplify the 
system (\ref{SYSGKODONE}) to the case of an isotropic conection $A^a_i=\delta^a_i{a}$, where $a=a(t)$ is a function only of time.

\subsection{Isotropic connection}

We now consider the individual terms.  Starting with the semiclassical term we 
have $\hbox{det}B=\vert{A}\vert^2$.  Moving on the the term first order in singularity we have

\begin{eqnarray}
\label{FIRSTSING}
\epsilon_{ijk}\epsilon^{abc}D^{kj}_{cb}B^i_e\Psi_{ae}
=\epsilon_{ijk}\epsilon^{abc}\epsilon^{kjl}\epsilon_{cbd}A^d_{l}B^i_{e}\Psi_{ae}\nonumber\\
(-2\delta^l_i)(-2\delta^a_d)A^d_{l}B^i_{e}\Psi_{ae}
=4C^a_{e}\Psi_{ae}=4(\delta^a_{e}a^3)\Psi_{ae}=4a^3\hbox{tr}\Psi
\end{eqnarray}

\indent
continuing on, we have

\begin{eqnarray}
\label{FIRSTSING1}
\epsilon_{ijk}\epsilon^{abc}B^k_{c}B^j_{e}\Psi_{be}
=\epsilon^{abc}\vert{B}\vert(B^{-1})^d_{i}\epsilon_{dec}\Psi_{be}\nonumber\\
=\bigl(\delta^a_{d}\delta^b_{e}-\delta^a_{e}\delta^b_{d}\bigr)
\vert{A}\vert^{2}A^d_{i}\vert{A}\vert^{-1}\Psi_{be}
=\vert{A}\vert\bigl(\delta^b_{e}A^a_{i}-\delta^a_{e}A^b_i\bigr)\Psi_{be}
\end{eqnarray}

\noindent
With a homogeneous and isotropic connection (\ref{FIRSTSING1}) yields

\begin{eqnarray}
\label{FIRSTSING11}
\vert{A}\vert\bigl(\delta^b_{e}A^a_{i}-\delta^a_{e}A^b_i\bigr)\Psi_{be}
=a^{4}\bigl(\delta^b_{e}\delta^a_{i}-\delta^a_{e}\delta^b_{i}\bigr)\Psi_{be}
=a^{4}\bigl(\delta^a_{i}\hbox{tr}\Psi-\Psi_{ia}\bigr)
\end{eqnarray}

Also, we have

\begin{eqnarray}
\label{FIRSTSING2}
\epsilon^{abc}\epsilon_{ijk}B^j_{f}B^k_{e}\Psi_{ce}\Psi_{bf}
=\epsilon^{abc}\vert{B}\vert(B^{-1})^d_{i}\epsilon_{dfe}\Psi_{ce}\Psi_{bf}\nonumber\\
=\vert{A}\vert^2\vert{A}\vert^{-1}A^d_{i}\epsilon^{abc}\epsilon_{dfe}\Psi_{ce}\Psi_{bf}
=\vert{A}\vert{A}^d_{i}\epsilon^{abc}\epsilon_{dfe}\Psi_{ce}\Psi_{bf}.
\end{eqnarray}

\noindent
With a homogeneous and isotropic connection (\ref{FIRSTSING2}) yields

\begin{eqnarray}
\label{FIRSTSING21}
\vert{A}\vert{A}^d_{i}\epsilon^{abc}\epsilon_{dfe}\Psi_{ce}\Psi_{bf}
=a^4\epsilon^{abc}\epsilon_{ife}\Psi_{ce}\Psi_{bf}
\end{eqnarray}

\noindent
So using (\ref{FIRSTSING}), (\ref{FIRSTSING11}) and (\ref{FIRSTSING21}), the system 
(\ref{SYSGKODONE}) for the case of a homogeneous, isotropic connection yields

\begin{eqnarray}
\label{SEM}
\hbox{det}B\bigl(GV\hbox{det}\Psi+Var\Psi\bigr)
+{{\pi^2} \over 2}=\vert{A}\vert^{2}\bigl(GV\hbox{det}\Psi+Var\Psi\bigr)
+{{\pi^2} \over 2}
\end{eqnarray}

\noindent
for the semiclassical term,

\begin{eqnarray}
\label{SEM1}
\epsilon_{ijk}\epsilon^{abc}D^{kj}_{cb}\Psi_{ae}B^i_{e}+
\epsilon_{ijk}\epsilon^{abc}{\partial \over {\partial{A}^a_i}}
\Bigl[B^k_{c}B^j_{e}\Psi_{be}+{V \over 4}B^k_{e}B^j_{f}\Psi_{ce}\Psi_{bf}\Bigr]
-{i \over {2G}}{{\partial\pi} \over \partial\phi}\nonumber\\=
4a^3\hbox{tr}\Psi+
\delta^i_a{\partial \over {\partial{a}}}
\Bigl[a^{4}\bigl(\delta^a_{i}\hbox{tr}\Psi-\Psi_{ia}\bigr)
+{V \over 4}a^4\epsilon^{abc}\epsilon_{ife}\Psi_{ce}\Psi_{bf}\Bigr]
-{i \over {2G}}{{\partial\pi} \over \partial\phi}\nonumber\\
=4a^3\hbox{tr}\Psi
+{\partial \over {\partial{a}}}
\Bigl[2a^{4}\hbox{tr}\Psi
+{V \over 4}a^{4}Var\Psi\Bigr]-{i \over {2G}}{{\partial\pi} \over \partial\phi}
\end{eqnarray}

\noindent
for the term first-order in singularity, and

\begin{eqnarray}
\label{SEM2}
{{GV} \over 6}{\partial \over {\partial{A^a_i}}}{\partial \over {\partial{A^b_j}}}
(\epsilon_{ijk}\epsilon^{abc}B^k_{e}\Psi_{ce})+36\nonumber\\
={{GV} \over 6}{{\partial^2} \over {\partial{a}^2}}\delta^i_{a}\delta^j_{b}
(\epsilon_{ijk}\epsilon^{abc}\vert{A}\vert(A^{-1})^k_{e}\Psi_{ce})+36
={{GV} \over 3}{{\partial^2} \over {\partial{a}^2}}\bigl(a^2\hbox{tr}\Psi\bigr)+36=0
\end{eqnarray}

\noindent
for the term second-order in singularity, the functional `Laplacian' term.

\subsection{Solution of the system}

We can now solve the system (\ref{SYSGKODONE}) for the homogeneous isotropic connection scenario.  Integrating the `functional Laplacian' equation (\ref{SEM2}), we can solve directly for $\hbox{tr}\Psi$ to obtain

\begin{eqnarray}
\label{TRACE}
\hbox{tr}\Psi=-{{54} \over {GV}}+{{b_1(\phi)} \over a}
+{{b_2(\phi)} \over {a^2}}
\end{eqnarray}

\noindent
where $b_1$ and $b_2$ are `constants' of integration with respect to $a$, which are in general functions of the matter field $\phi$.  Prior to substitution into (\ref{SEM1}) we invoke the mixed partials condition in order to eliminate the matter momentum $\pi$ from the system.

\begin{eqnarray}
\label{PARTIALSS}
{{\partial\pi} \over {\partial{A}^a_i}}
=-{i \over G}B^i_{e}{{\partial\Psi_{ae}} \over {\partial\phi}}.
\end{eqnarray}

\noindent
For a homogeneous and isotropic connection this reads

\begin{eqnarray}
\label{PARTIALSS1}
\delta^i_{a}{{\partial\pi} \over {\partial{a}}}
=-{i \over G}a^{2}{{\partial\Psi_{ai}} \over {\partial\phi}}.
\end{eqnarray}

\noindent
Taking the trace of (\ref{PARTIALSS1}), this yields

\begin{eqnarray}
\label{PARTIALSS2}
{{\partial\pi} \over {\partial{a}}}
=-{i \over {3G}}a^{2}{\partial \over {\partial\phi}}\hbox{tr}\Psi
\end{eqnarray}

\noindent
which, upon integration with respect to $a$ yields

\begin{eqnarray}
\label{PARTIALSS3}
\pi=f(\phi)-{i \over {3G}}\int{da}~a^{2}{\partial \over {\partial\phi}}\hbox{tr}\Psi
\end{eqnarray}

\noindent
where $f(\phi)$ is an arbitrary function of the matter field $\phi$, which acts as a `constant' of integration with respect to $a$, as well as a functional boundary condition on the semiclassical matter momentum.  The function $f$ can be thought of as the matter momentum of the quantum state in the absence of gravitational interaction.  Substituting (\ref{TRACE}) into (\ref{PARTIALSS3}), we obtain that 

\begin{eqnarray}
\label{PARTIALSS4}
\pi=\pi(a,\phi)=f(\phi)+{i \over {3G}}\Bigl[-54{{V^{\prime}} \over {GV^2}}
+{1 \over 2}b_{1}^{\prime}(\phi)a^2
+b_{2}^{\prime}(\phi)a\Bigr]
\end{eqnarray}

\noindent
where a prime denotes differentiation with respect to the matter field $\phi$.  This expression for the matter momentum can be substituted into the original system to solve for the invariants $Var\Psi$ and $\hbox{det}\Psi$, which can in turn be used to solve for the CDJ matrix elements.  But as we shall see this is not required in order to determine the generalized Kodama state $\Psi_{GKod}$ for the homogeneous and isotropic case.

\subsection{Determination of the generalized Kodama state.}

To determine the generalized Kodama state we must evaluate 

\begin{equation}
\label{KOMS}
I=\int_{M}\Bigl((\hbar{G})^{-1}\Psi_{ae}B^i_{a}\dot{A}^e_{i}
+{i \over \hbar}\pi\dot{\phi}\Bigr)
\end{equation}

\noindent
for the case of the homogeneous and isotropic connection.  This yields

\begin{equation}
\label{KOMSA}
I=l^3\int{dt}((\hbar{G})^{-1}\dot{a}a^2\hbox{tr}\Psi
+{i \over \hbar}\pi\dot{\phi}\Bigr)
\end{equation}

\noindent
where $l$ is the characteristic length scale of the spatial manifold $\Sigma$ comprising the universe, making use of minisuperspace.  This leaves remaining an integral over time.  Making use of the relations $dt\dot{a}=da$ and $dt\dot{\phi}=d\phi$ and the fact that the connection $a$ and the matter field $\phi$ are independent dynamic variables, this allows us to deparametrize in accordance with the arguments presented in appendix A and section 11 of \cite{EYO} any detailed time dependence of the state.

\begin{equation}
\label{KOMSA1}
I=l^3\int\Bigl((\hbar{G})^{-1}da~a^2\hbox{tr}\Psi
+{i \over \hbar}\pi{d}\phi\Bigr)
\end{equation}

\noindent
where now the integral is over the functional space of fields.  This circumvents the problem of time in that the state is transparent to the mechanism of its evolution from an initial to a final hypersurface.  There are two contributions to (\ref{KOMSA1}).  First there is the gravitational sector contribution, given by

\begin{eqnarray}
\label{GRA}
\int(\hbar{G^2})^{-1}da~a^2\hbox{tr}\Psi
=(\hbar{G^2})^{-1}\int{da}~a^2
\Bigl(-{{54} \over V}+{{b_1(\phi)} \over a}+{{b_2(\phi)} \over {a^2}}\Bigr)\nonumber\\
=(\hbar{G^2})^{-1}\Bigl(-{{18} \over V}a^3+{1 \over 2}b_{1}(\phi)a^2+b_{2}(\phi)a\Bigr).
\end{eqnarray}

\noindent
Then there is a contribution form the matter sector, given by

\begin{eqnarray}
\label{GRA1}
{i \over \hbar}\int\pi{d}\phi
={i \over \hbar}\int{d}\phi
\Bigl(f(\phi)-{i \over {3G}}{\partial \over {\partial\phi}}
\Bigl[-{{18} \over {GV}}a^3+{1 \over 2}b_{1}(\phi)a^2+b_{2}(\phi)a\Bigr]\Bigr)\nonumber\\
={i \over \hbar}\int{d}\phi{f}(\phi)
+(3G\hbar)^{-1}\Bigl[-{{18} \over {GV}}a^3+{1 \over 2}b_{1}(\phi)a^2+b_{2}(\phi)a\Bigr].
\end{eqnarray}

\noindent
So overall, the generalized Kodama state is given by

\begin{eqnarray}
\label{KODAMA}
\Psi_{GKod}[a,\phi]
=\hbox{exp}\Bigl[{{il^3} \over \hbar}\int{d}\phi{f}(\phi)\Bigr]
\hbox{exp}\Bigl[{{4l^3} \over 3}(\hbar{G^2})^{-1}
\Bigl(-{{18} \over V}a^3+{1 \over 2}b_{1}(\phi)a^2+b_{2}(\phi)a\Bigr)\Bigr]
\end{eqnarray}

\noindent
Equation (\ref{KODAMA}) is the solution to the quantized Hamiltonian constraint from the vantage point of minisuperspace reduction after quantization.  It contains three main degrees of freedom relative to the pure Kodama state for the corresponding model devoid of the matter field.  There are the two `constants' of integration $b_1$ and $b_2$, which can be chosen judiciously to enforce certain desirable properties upon the norm of $\Psi_{GKod}$.  For example, it may be desirable for the wavefunction of the universe to not have support for trivial and/or infinite values of the connection.  Such criteria fix the values and signs of these functions.  The third main degree of freedom is in the function $f(\phi)$.  This essentially determines a basis of eigenfunctions in which to expand the matter momentum.  For example, in the slow-roll basis one may impose the restriction that the matter sector of the generalized Kodama state be expressible in terms of eigenvalues of the slow-roll parameter $r$, such that

\begin{eqnarray}
r={{\pi^2} \over {2V(\phi)}}
\end{eqnarray}

\noindent
as may be compared to \cite{INFL}.  In this case we have for the matter sector

\begin{eqnarray}
{i \over \hbar}\int_{M}\pi{d\phi}
={{il^3} \over \hbar}\sqrt{2r}\int{V}^{1/2}d\phi
={{2il^3} \over {3\hbar}}\sqrt{2r}V^{3/2}(\phi),
\end{eqnarray}

\noindent
where the value of $\phi$ is evaluated on the final hypersurface.  This as well highlights the noncommutativity of functional integration with time integration, a property which we argue here and in \cite{EYO} extends to the full theory.

\subsection{Physical interpretation of the generalized Kodama states}

\noindent
To see the connection between these results and the formalism set up in \cite{EYO}, it will be instructive to analyze some mechanisms by which the existence of generalized Kodama states may possibly manifest themselves in an appropriate semiclassical limit.  Assume, for a toy model, that the matter content of the observational universe that we live in today is governed in the observable limit by the laws of quantum field theory on Minkowski spacetime.  In this description, the effects of the perturbative regime of quantized gravity in metric variables are very weak so as to be negligible in the effective theories at existing accelerator energy scales  \cite{ORDER1},\cite{WEIN1}.  But we argue for the possiblity that the nonperturbative effects of quantum gravity could produce clearly distinguishable effects, even in the semiclassical limit.\par
\indent
It appears naively in (\ref{KODAMA}) that there are no restrictions on $f(\phi)$, which is a direct result of the nonperturbative canonical quantization procedure outlined in \cite{EYO} and $V(\phi)$, a (naively) freely specifiable input from the matter sector into the process.  However, one can acquire a physical interpretation for how these functional parameters of the field $\phi$ may be related by considering the semiclassical limit.  If one makes the definition 

\begin{eqnarray}
\label{CONDITION4}
\Theta_f(\phi)=\int^{\phi}_{\phi_0}{d\varphi}f(\varphi)
\end{eqnarray}

\noindent
then one can write the generalized Kodama state (\ref{KODAMA}), taking $b_1=b_2=0$ for simplicity, as

\begin{eqnarray}
\label{KODAMA1}
\Psi_{GKod_f}[a,\phi]
=e^{i{{l^3} \over \hbar}\Theta_f(\phi)}
\hbox{exp}\Bigl[-{24l^3}(\hbar{G^2})^{-1}
\Bigl({{a^3} \over {V(\phi)}}\Bigr)\Bigr].
\end{eqnarray}

\noindent
Equation (\ref{KODAMA1}) consists of a gravitational sector and a matter sector labeled by the function $f$.  For a physical interpretation of the relationship between $\Theta_f$ and $V$ it is instructive to consider the limit $a\rightarrow{0}$.  Configurations for which $a=0$ correspond to a degenerate $B^i_a$, which are not allowed when coupling quantum gravity to matter fields by our method \cite{EYO}.  This can be seen from the first equation of (\ref{SYSGKODONE}), which applied in the case of the homogeneous isotropic connection $a$, yields

\begin{eqnarray}
\label{CONDITION}
Var\Psi+GV\hbox{det}\Psi={{G\Omega_0} \over {a^2}}.
\end{eqnarray}

\noindent
Since $\hbox{det}B=a^2$ in this case, one must either (i) require $\Omega_0=0$, which means that any matter fields $\phi$ must as well vanish in this limit, or (ii) realize that the method is being applied beyond its realm of validity and provide some observational input from physics to cover this case.  Taking $a=0$ 
in (\ref{KODAMA1}) collapses the generalized Kodama state into just the matter contribution, which can be interpreted in this case as $\Psi_{GKod}$ in the absence of gravity.  So the fundamentally relevant question then becomes what physics determines the wavefunction of the universe in this case.\par
\indent
If one assumes that the governing laws of the universe in the absence of gravity are dictated by quantum field theory on Minkowski spacetime, then one can interpret 
$\Theta$ in (\ref{CONDITION4}) as the phase of the corresponding wavefunction in Minkowski spacetime.  Of course, as long as there is matter content in the universe, there should be gravity by virtue of the Einstein's equations.  Hence there should always by definition exist a backreaction between the matter and the gravitational sectors in a viable description.  In the toy model considered in this paper, the governing equation in the gravity-free limit would be the Schr\"odinger equation for the Klein--Gordon field in minisuperspace.  This arises from the Hamiltonian $H_{matter}=\pi^2/2+V(\phi)$.  The corresponding Schr\"odinger equation is given by

\begin{eqnarray}
\label{KGSEEEEE}
\Bigl[-{{\hbar^2} \over 2}{\delta \over {\delta\phi}}{\delta \over {\delta\phi}}+V(\phi)\Bigr]\chi(\phi)=E\chi(\phi).
\end{eqnarray}

\noindent
where $E$ is the energy eigenvalue of the state.  Equation (\ref{KGSEEEEE}) can be solved exactly for the eigenstates of the system in a few cases.  The important point is that when one makes the identification

\begin{eqnarray}
\label{NOGRAV}
\chi(\phi)=e^{{{il^3} \over \hbar}\Theta_f(\phi)},
\end{eqnarray}

\noindent
then the relationship between $\Theta$ and $V$ in (\ref{KODAMA1}) is fixed by imposition of the proper semiclassical limit.  From this perspective $\Psi_{GKod}$ provides a wavefunction for the universe that exhibits a definite semiclassical limit both (i) for pure gravity with $\Lambda$ term in the absence of 
matter\footnote{$\Psi_{Kod}$ corresponding to DeSitter spacetime \cite{POSLAMB},\cite{KOD}}, and (ii) for pure matter in the absence of 
gravity\footnote{$\chi$ corresponding to quantum field theory on Minkowski spacetime}.  This applies to the full theory as well as to minisuperspace, irrespective of the matter model being considered.\par
\indent
One solves (\ref{KGSEEEEE}) for the orthonormal eigenstates of the system.  Therefore the generalized Kodama state $\Psi_{Kod}$ acquires the label of these eigenstates.  The crux of the phenomenological applications of our method is that one first assesses whether the matter sector is in the Lorentzian (classical) region of configuration 
space ($\Theta$ real) or the Euclidean region (tunneling) region ($\Theta$ imaginary).  The full-blown $\Psi_{GKod}$ then extrapolates this semiclassical input to the regime $a\neq{0}$ which corresponds to the presence of quantized gravity coupled to the quantized matter fields.  Note that although the eigenstates $\chi_n$ may be orthogonal in the gravity-free case $(a=0)$, they in general will not be orthonormal within the complete $\Psi_{Kod}$ for $a\neq{0}$ due to the $\phi$ dependence of $V(\phi)$ which acts as a cosmological constant for the gravitational sector.  This might lead to experimentally observable effects that our model can distinguish\footnote{For example transitions between quantum states $\chi_n$ that are shown to be forbidden in the Minkowski limit of spacetime}.\par
\indent
For example, for a massive Klein--Gordon field $V=(1/2)m^2\phi^2$.  Equation (\ref{KGSEEEEE}) corresponds to a simple harmonic oscillator of `unit' frequency.  The eigenstates form an orthonormal basis in Minkowski spacetime, but lead to a $\Psi_{GKod}$, assuming that the cosmological constant is due entirely to the presence of the matter field and defining $\rho=\sqrt{{ml^3}/\hbar}\phi$, of

\begin{eqnarray}
\label{NOGRAV1}
\Psi_{GKod_n}(a,\rho)={1 \over {\sqrt{2^nn!}}}H_{n}(\rho)e^{-{{\rho^2} \over 2}}
\hbox{exp}\Bigl[-{48l^6}(\hbar^2{G^2}m)^{-1}\Bigl({{a^3} \over {\rho^2}}\Bigr)\Bigr].
\end{eqnarray}

\noindent
Equation (\ref{NOGRAV1}) corresponds to a tunneling matter sector coupled to gravity.  Note in (\ref{NOGRAV1}) that $\Psi_{GKod_n}$ vanishes for $\phi=0$.  In order to produce the correct semiclassical limit of DeSitter spacetime for $\phi=0$, one must augment the matter potential by the cosmological constant $\Lambda$ as 
in $V(\phi)={\Lambda \over G}+(m/2)\phi^2$.  This provides some input from the gravitational into the nongravitational sector which is directly observable\footnote{There is vast cosmological evidence that points to a small but positive $\Lambda$ \cite{COSMO}.}, which one can see by substituting the new potential $V$ into 
(\ref{KGSEEEEE}).  The governing equation for the nongravitational limit of quantum field theory on Minkowski spacetime then inherits the parameter $\Lambda/G$ from quantum gravity, which affects the eigenstates $\chi_n$ and their energy eigenvalues: thus even in the absence of gravity, there is gravity!\par
\indent
For the more general case it may not be possible to solve (\ref{KGSEEEEE}) exactly, but one may use the WKB approximation to any desired order obtain the semiclassical limit for the matter sector.  For example, the first-order approximation to $\Psi_{GKod}$ would be given by

\begin{eqnarray}
\label{NOGRAV}
\Psi_{GKod}(a,\phi)\sim{e}^{{i \over \hbar}\int^{\phi}_{\phi_0}{d\varphi}\sqrt{2(E-V(\varphi))}}
\hbox{exp}\Bigl[-{24l^3}(\hbar{G^2})^{-1}
\Bigl({{a^3} \over {V(\phi)}}\Bigr)\Bigr]
\end{eqnarray}

\noindent
Equation (\ref{NOGRAV}) clearly shows how $f$ and $\Theta$ can be constrained.  There are additional constraints on these functions based on considerations of normalizability of $\Psi_{GKod}$, as well as theoretical and observational evidence for classical versus tunneling configurations which it would be interesting to illustrate.  We will save such considerations for future work.

\section{Semiclassical solution to the Hamiltonian constraint}

We now show a novel method for solving the Hamiltonian constraint at the semiclassical level, in order to provide some more results.  We expect from paper one that there is an infinite two-parameter family of solutions, when account is not taken of the equations $q_1=q_2=0$ arising from the quantum terms.  We will consider a subset of this two-parameter family, namley the case of a homogeneous and isotropic CDJ matrix $\Psi_{ae}$.  The difference from the case considered above is that now the invariants of the CDJ matrix, $\hbox{tr}\Psi$, $Var\Psi$ and $\hbox{det}\Psi$ are no longer independent variables to be solved for.  These are determined by $\hbox{tr}\Psi$.\par
\indent
When quantum gravity in Ashtekar variables with cosmological constant $\Lambda$ is coupled to matter fields, an interesting effect happens.  The SQC remains intact for the
kinematic constraints due to the fact that the kinetic terms of the matter fields are generally no higher than second-order in matter momenta, and the corresponding Lie algebra of
constraints still closes on these constraints \cite{ASH1}.  So the problematic constraint is the dynamic one, the Hamiltonian constraint.  The Hamiltonian constraint is what 
distinguishes one diffeomorphism-gauge invariant theory from another, and leads to an ambiguity amongst quantum theories of gravity arising from the same classical theory.\par
\indent
Revisiting the case of a Klein-Gordon scalar field $\phi$ with self-interaction potential 
$V(\phi)$ and conjugate momentum $\pi$.  The classical Hamiltonian constraint 
reads \cite{ASH}

\begin{equation}
\label{PAP8}
 H={\epsilon^{abc}}{\epsilon_{ijk}}\widetilde{\sigma}^i_a\widetilde{\sigma}^j_{b}B^k_c
 +{1 \over 6}{\epsilon^{abc}}{\epsilon_{ijk}}\widetilde{\sigma}^i_a\widetilde{\sigma}^j_b\widetilde{\sigma}^k_{c}V(\phi)
 +{1 \over 2}{\pi}^2
 +{1 \over 2}\delta^{ab}\widetilde{\sigma}^i_a\widetilde{\sigma}^j_b{\partial_i{\phi}}{\partial_j{\phi}}=0.
 \end{equation}

\noindent 
Making use of the identity

\begin{equation}
{{{\epsilon^{abc}}{\epsilon_{ijk}}{\widetilde{\sigma}}^i_a{\widetilde{\sigma}}^j_b{\widetilde{\sigma}}^k_c}\over {6\hbox{det}\widetilde{\sigma}}}=1,
\end{equation}
  
\noindent
we multiply the ${{\pi^2} \over 2}$ term of (\ref{PAP8}) by $1$ in this form and rewrite the semiclassical part of the Hamiltonian constraint in the form

\begin{equation}
\widetilde{\sigma}^i_a\widetilde{\sigma}^j_b\bigl[A^{ab}_{ij}+S^{ab}_{ij}\bigr]=0,
\end{equation}

\noindent
where $A^{ab}_{ij}$ is antisymmetric in its ab and ij indices, given by

\begin{equation}
A^{ab}_{ij}={\epsilon^{abc}}{\epsilon_{ijk}}\Bigl[B^k_c
+\widetilde{\sigma}^k_c\Bigl({1 \over 6}V(\phi)
+{{\pi}^2 \over {12\hbox{det}{\widetilde{\sigma}}}}\Bigr)\Bigr],
\end{equation}

\noindent
and $S^{ab}_{ij}$ is symmetric in its indices, given by

\begin{equation}
S^{ab}_{ij}={1 \over 2}{\delta^{ab}}{\partial_i{\phi}}{\partial_j{\phi}}.
\end{equation}

\noindent
A possible nontrivial solution for nonzero $\widetilde{\sigma}^i_a$ is to require $A^{ab}_{ij}=0$ and $S^{ab}_{ij}=0$ $\forall~i,j,a,b$, which is the condition that the 
symmetric and antisymmetric parts of a vanishing tensor must separately vanish.  This Ansatz can be motivated by analogy to the pure gravity case, for which there is no 
$S^{ab}_{ij}$.  The condition $S^{ab}_{ij}=0$ would imply that $\partial_{i}\phi=0~\forall{i}$.  This may appear to be restrictive, but let us nevertheless proceed as though 
$S^{ab}_{ij}$ can be neglected and examine the consequences.\par
\indent
So we must have $A^{ab}_{ij}=0$.  A nontrivial solution, in direct analogy to the steps leading to the pure Kodama state, is that

\begin{equation}
\label{PAP9}
-B^k_c=\widetilde{\sigma}^k_c\Bigl({1 \over 6}V(\phi)+{{\pi}^2 \over {12\hbox{det}{\widetilde{\sigma}}}}\Bigr)~\forall~k,c.
\end{equation}

\noindent
Taking the determinant of both sides of (\ref{PAP9}) and rearranging, we have

\begin{equation}
\label{PAP10}
{{\hbox{det}B} \over {\hbox{det}{\widetilde{\sigma}}}}+\Bigl({1 \over 6}V(\phi)+{{\pi}^2 \over {12\hbox{det}{\widetilde{\sigma}}}}\Bigr)^3=0
\end{equation}

\noindent
which is a cubic equation in $\hbox{det}\widetilde\sigma$.  We illustrate in detail the method of closed-form solution in the appendix, for completeness.  The polynomial structure of the constraints expressed in the Ashtekar variables is the major simplification of general relativity which makes this possible.  Note that the existence of a 
meaningful solution rests on the nondegeneracy of the spatial 3-metric ($\hbox{det}\widetilde\sigma\neq{0}$).  The restriction to nondegenerate 3-metrics $h_{ij}$ when matter is included may perhaps be interpreted as the requirement that the signature of the metric be 
fixed.\par
\indent
A solution to (\ref{PAP10}) upon substitution back into (\ref{PAP9}) leads to the condition

\begin{equation}
\widetilde{\sigma}^i_a(x)=-(\Lambda_{eff})^{-1}B^i_a(x)~~\forall~x,
\end{equation}

\noindent
which can be interpreted as a 'generalized' self-duality relation of the Ashtekar electric to magnetic field, with a generalized field-dependent 'cosmological constant'
$\Lambda_{eff}$, given by

\begin{equation}
\Lambda_{eff}=4\Bigl({{\hbox{det}B} \over {\pi^2}}\Bigr)^{1/2}
\hbox{sinh}\bigl[(1/3)\hbox{sinh}^{-1}\bigl({V \over 8}\Bigl({{\pi^2} \over {\hbox{det}B}}\Bigr)^{1/2}\bigr)\bigr].
\end{equation}

\noindent
This is analogous to the phenomenon of dispersion in Maxwell theory for electromagnetic waves propagating in a material medium.  Due to the presence of the matter field the
Ashtekar electric field has become distorted from its pure gravity value 
$\widetilde{\sigma}^i_a=-6\Lambda^{-1}B^i_a$, or alternatively, rescaled itself in an attempt to
restore a broken semiclassical-quantum correspondence, broken due to the presence of the inhomogeneous matter terms.\par
\indent

\bigskip
\section{Semiclassical orbits of spacetime}
\par
\medskip
\indent
Let us examine the semiclassical orbits of the spacetime resulting from the constraint.  We will use the minisuperspace approximation for simplicity, but
nevertheless should be able to derive some nonperturbative analytical information from the solution to the constraint.  We would like to determine possible nonperturbative 
effects upon the mechanisms of inflation.  It is known that the pure Kodama state possesses as one of its semiclassical orbits the deSitter spacetime \cite{POSLAMB}, which 
corresponds to an exponentially inflating universe with metric

\begin{equation}
ds^2=dt^2+e^{\sqrt{\Lambda \over 3}t}\delta_{ij}dx^{i}dx^j.
\end{equation}

We would like to determine what type of spacetimes the generalized Kodama semiclassical state predicts.  In the case when the matter field is a Klein-Gordon scalar 
field we should hope to attain a closer analysis of inflation which can be compared with observational data.  Starting from the Hamiltonian in the zero shift $(N^i=0)$ gauge, for 
simplicity,

\begin{equation}
 H=\int_{\Sigma}i\underline{N}(\epsilon^{abc}\epsilon_{ijk}\widetilde{\sigma}^i_{a}\widetilde{\sigma}^j_{b}B^k_c
 +{1 \over 6}{\epsilon^{abc}}{\epsilon_{ijk}}\widetilde{\sigma}^i_a\widetilde{\sigma}^j_b\widetilde{\sigma}^k_{c}V(\phi)
 +{1 \over 2}{\pi}^2),
 \end{equation}

\noindent
we can write the Hamiltonian equations of motion.

\begin{eqnarray}
\dot{A}^a_i={i \over 2}\underline{N}{\epsilon^{abc}}{\epsilon_{ijk}}(\widetilde{\sigma}^j_{b}\widetilde{\sigma}^k_{c}V(\phi)+\widetilde{\sigma}^j_{b}B^k_c)\nonumber\\
\dot{\widetilde\sigma}^i_a=-i\epsilon^{ilm}D_{l}\bigl(\underline{N}(\hbox{det}\sigma)(\widetilde\sigma^{-1})^c_m)\delta_{ac}\nonumber\\
\dot\phi=\pi\nonumber\\
\dot\pi=i\underline{N}V^{\prime}\hbox{det}\widetilde\sigma
\end{eqnarray}

\noindent
where $V^{\prime}=(\partial{V}/\partial\phi)$.  We will use an isotropic spatially homogeneous, imaginary connection as in \cite{POSLAMB}, hence

\begin{equation}
A^a_i=if(t)\delta^a_i\longrightarrow{B}^i_a=-f^2(t)\delta^i_a,
\end{equation}

\noindent
where $f(t)$ is a spatially homogeneous function of time.  We now apply the generalized self-duality condition arising from the Hamiltonian constraint
$\widetilde{\sigma}^i_a=-(\Lambda_{eff})^{-1}B^i_a$, using

\begin{equation}
\Lambda_{eff}=\Bigl({{4if^3} \over {\pi}}\Bigr)T_{1/3}\Bigl[{{iV\pi} \over {8f^3}}\Bigr]
=4\Bigl({{\hbox{det}B} \over {\pi^2}}\Bigr)^{1/2}
\hbox{sinh}\bigl[(1/3)\hbox{sinh}^{-1}\bigl({V \over 8}\Bigl({{\pi^2} \over {\hbox{det}B}}\Bigr)^{1/2}\bigr)\bigr]
\end{equation}

\noindent
We will also need the relations

\begin{equation}
\widetilde{\sigma}^i_a=-{{B^i_a} \over {\Lambda_{eff}}}={{f^2} \over {\Lambda_{eff}}}\delta^i_a\longrightarrow\hbox{det}\sigma
={{f^6} \over {(\Lambda_{eff})^3}}.
\end{equation}

\noindent
We also make the observation as in \cite{POSLAMB} that since $\underline{N}$ is a scalar density it goes as 
$\hbox{det}^{-1/2}\widetilde\sigma\propto{f}^{-3}(\Lambda_{eff})^{3/2}$.  Numerical constants and such factors of $i$ can always be absorbed into the definition of $\underline{N}$.  Substituting into 
the equations of motion we have, for $f$

\begin{equation}
\dot{f}=\Bigl({{(\Lambda_{eff})^3} \over {f^6}}\Bigr)^{1/2}\Bigl({{f^2} \over {\Lambda_{eff}}}\Bigr)^2\bigl[V+\Lambda_{eff}\bigr]
=f(\Lambda_{eff})^{-1/2}\bigl[V+\Lambda_{eff}\bigr].
\end{equation}

\noindent
We perform some algebraic manipulations to put it into the form

\begin{equation}
\dot{f}=f\sqrt{V}\Bigl[\Bigl({{\Lambda_{eff}} \over V}\Bigr)^{1/2}+\Bigl({{\Lambda_{eff}} \over V}\Bigr)^{-1/2}\Bigr].
\end{equation}

So we have, for the Ashtekar connection,

\begin{equation}
{d \over {dt}}\hbox{ln}f=2\sqrt{V}\hbox{cos}\Bigl[{1 \over 2}\hbox{ln}\Bigl({V \over {\Lambda_{eff}}}\Bigr)\Bigr]
=2\sqrt{V}\hbox{cos}\Bigl[{1 \over 2}\hbox{ln}\Bigl({{T_{1/3}(\alpha)} \over {2\alpha}}\Bigr)\Bigr],
\end{equation}

\noindent
where we have defined 

\begin{equation}
\alpha={{iV\pi} \over {8f^3}}.
\end{equation}

The following relations will be useful in nonperturbatively deducing the dynamics for various regimes of curvature.

\begin{equation}
\hbox{lim}_{\alpha\rightarrow{0}}\Bigl({{T_{1/3}(\alpha)} \over \alpha}\Bigr)={1 \over 3};~~~
\hbox{lim}_{\alpha\rightarrow\infty}\Bigl({{T_{1/3}(\alpha)} \over \alpha}\Bigr)(2\alpha)^{-2/3}\sim\alpha^{-2/3}.
\end{equation}

\noindent
where we have defined the Tchebyshev polynomial

\begin{eqnarray}
T_{1/3}(x)={1 \over x}\hbox{sinh}\bigl[(1/3)\hbox{sinh}^{-1}x\bigr]
\end{eqnarray}

Note that $\alpha\rightarrow{0}$ corresponds to an infinite curvature singularity, covering perhaps the initial stages of the universe starting from the big bang scenario 
and that $\alpha\rightarrow\infty$ corresponds to degenerate curvature, perhaps closer to the universe of today (nearly flat spacetime).\par
\indent
Deriving the equation of motion for the scalar field $\phi$, we have

\begin{equation}
{{d^2} \over {dt^2}}\phi=i\Bigl({{f^6} \over {(\Lambda_{eff})^3}}\Bigr)^{1/2}{{dV} \over {d\phi}}
=-if^3\Bigl({\alpha \over {T_{1/3}(\alpha)}}\Bigr)^{-3/2}V^{-3/2}{{dV} \over {d\phi}}.
\end{equation}

\noindent
Or we have

\begin{equation}
\label{PAP11}
{{d^2} \over {dt^2}}\phi=-2if^3\Bigl({{d\sqrt{V}} \over {d\phi}}\Bigr)\Bigl({\alpha \over {T_{1/3}(\alpha)}}\Bigr)^{-3/2}
\end{equation}

\noindent
This can also be written, multiplying both sides of (\ref{PAP11}) by $\dot\phi$,

\begin{equation}
{d \over {dt}}\bigl({{\pi^2} \over 2}\bigr)
=-2if^3\Bigl({{d\sqrt{V}} \over {dt}}\Bigr)\Bigl({\alpha \over {T_{1/3}(\alpha)}}\Bigr)^{-3/2}.
\end{equation}

Rewriting the equations of motion, for completeness, we have the two simultaneous classical equations of motion for the Ashtekar/Klein-Gordon minisuperspace model, linked by
$\Lambda_{eff}$ as determined by the Hamiltonian constraint,

\begin{equation}
{d \over {dt}}\hbox{ln}f=2\sqrt{V}\hbox{cos}\Bigl[{1 \over 2}\hbox{ln}\Bigl({{T_{1/3}(\alpha)} \over {2\alpha}}\Bigr)\Bigr];~~~
{{d^2} \over {dt^2}}\phi=-2if^3\Bigl({{d\sqrt{V}} \over {d\phi}}\Bigr)\Bigl({\alpha \over {T_{1/3}(\alpha)}}\Bigr)^{-3/2}.
\end{equation}

\noindent
The equation of motion for $f$ can be integrated to yield

\begin{equation}
f(t)=f(0)\hbox{exp}\bigl[\int^t_{t_0}dt~2\sqrt{V}\hbox{cos}\Bigl[{1 \over 2}\hbox{ln}\Bigl({{T_{1/3}(\alpha)} \over {2\alpha}}\Bigr)\Bigr]\biggr],
\end{equation}

\noindent
and substituted into the second equation of motion to determine the evolution of the scalar field $\phi$, and also used to determine the evolution of the spatial 3-metric
via the identification

\begin{equation}
h^{ij}=\hbox{det}^{-1}\widetilde\sigma\widetilde{\sigma}^i_a\widetilde{\sigma}^j_a=\Bigl({{\Lambda_{eff}} \over {f^2}}\Bigr)\delta^{ij}.
\end{equation}

\noindent
We are now ready to examine some regimes of interest.\par
\noindent

\subsection{First regime: Big bang singularity}

In this regime we have $(\hbox{det}B\rightarrow\infty)$ such that 
$\Lambda_{eff}\rightarrow{V \over 6}$ and the resulting evolution

\begin{equation}
f(t)=f(0)\hbox{exp}\Bigl[k\int^t_{t_0}dt~\sqrt{V}\Bigr],
\end{equation}

\noindent
where $k=2\hbox{cos}(\hbox{ln}\sqrt{6})$, which leads to a spacetime 3-metric

\begin{equation}
\label{PAP12}
h^{ij}=-\Bigl({V \over {6\hbox{det}^{1/3}B_0}}\hbox{exp}\Bigl[-2k\int^t_{t_0}dt~\sqrt{V}\Bigr]\Bigr)\delta^{ij},
\end{equation}

\noindent
where $B_0$ is the value of the Ashtekar curvature at the singularity at $t=0$.  Comparing this with $e^{\sqrt{\Lambda \over 3}t}$ from ref\cite{POSLAMB} one may naively 
conclude that this is an exponentially decaying rather than inflating spacetime.  However, the factor of $V$ multiplying the exponential in (\ref{PAP12}) causes an initial 
expansion for increasing positive $V$, which then begins recontraction at some critical time.  Whichever phase we are in today, if this is an accurate model, would have to be 
prior to that critical time.  Note that the negative sign on $h_{ij}$ would imply a Lorentzian signature for that spacetime.\par
\indent
However, if we make the replacement $\sqrt{V}\rightarrow-\sqrt{V}$ which amounts to taking the negative vice positive square root corresponding to $V^{1/2}\equiv\pm\sqrt{V}$, then we do get an exponentially inflating universe of Euclidean signature, inflating faster than the 
exponential rate.  The main result from this analysis is that whether or not the universe underwent a period of inflation fixes the signature of the spacetime.  To ascertain 
how the initial period of inflation progresses we must examine the equation of motion for $\phi$.  Note, also, that for negative potentials $V(\phi)$ the metric exhibits 
oscillatory behavior with Euclidean signature.  The equation of motion for $\phi$ is 

\begin{equation}
{{d^2} \over {dt^2}}\phi=-2i(3)^{-3/2}\Bigl({{d\sqrt{V}} \over {d\phi}}\Bigr)\hbox{exp}\Bigl[3k\int^t_{t_0}dt~\sqrt{V}\Bigr]
\end{equation}

\noindent
This can be examined for various forms of the potential.  For instance, for a mass squared term $V(\phi)={1 \over 2}m^2\phi^2$ this yields, upon taking the logarithm and
differentiating with respect to $t$, the equation

\begin{equation}
{{d^3} \over {dt^3}}\phi={{3km} \over {\sqrt{2}}}\phi{{d^2} \over {dt^2}}\phi,
\end{equation}

\noindent
which can be solved, if necessary, by numerical methods or even by Taylor expansion by taking repeated derivatives.  This should be the case for any potential $V(\phi)$.
Using a power series Ansatz of the form $\phi(t)=a_0+a_{1}t+a_2{t^2}+a_{0}t^3+...$, to find the short time behavior close to the initial singularity, it suffices to know the 
initial field $a_0=\phi(0)$, its initial velocity $a_1=\pi(0)$ and its initial acceleration $a_{2}=\dot\pi(0)$ to find the next important term, yielding

\begin{equation}
\phi(t)=a_0+a_{1}t+a_{2}t^2+{{km} \over {4\sqrt{2}}}a_2\bigl(a_0^2+a_1\bigr)t^3+...
\end{equation}

\noindent
for a potential of the form $V(\phi)={\Lambda \over {4!}}\phi^4$ the differential equation becomes

\begin{equation}
\phi{{d^3} \over {dt^3}}\phi={\sqrt{\Lambda \over {4!}}}(2\dot{\phi}+3k\phi^3){{d^2} \over {dt^2}}\phi
\end{equation}

\noindent
with a power series solution, at short times, of

\begin{equation}
\phi(t)=a_0+a_{1}t+a_{2}t^2+\sqrt{{\Lambda \over {4!}}}a_{2}\bigl({1 \over 3}a_{1}a_0^{-1}+a_0^{2}k\bigr)t^3+...
\end{equation}

It would be an interesting exercise to compare different potentials in the 3rd-order term of the expansion for within short times (the Planck time or before) of the singularity, 
assuming there was a big bang scenario, to obtain any experimentally verifiable effects delineating qualitative behavior of the resulting spacetime.

\par
\noindent

\subsection{Second regime: Region of possible signature change}

This is the opposite extreme $(\hbox{det}B\rightarrow{0})$.  In this limit we have $\Lambda_{eff}\rightarrow{0}$, but we would like to capture the specific functional dependence in order to make some meaningful predictions.  In this regime we have 

\begin{equation}
\Lambda_{eff}=\Bigl({{\pi^2} \over V}\Bigr)^{-1/3}f^2\longrightarrow\widetilde{\sigma}^i_a=\Bigl({{\pi^2} \over V}\Bigr)^{1/3}\delta^i_a
\end{equation}

\noindent
or that the metric depends not upon the curvature, but upon a ratio $r=(\pi^2/V)$, the ratio of the Klein-Gordon field kinetic energy to its potential energy.  This is known as 
the slow-roll parameter, which by \cite{INFL} is a measure of the departure of the dynamics from inflation.  The three-metric is then given by $h^{ij}=r^{1/3}\delta^{ij}$, or is 
constant in time for all configurations with the same value of $r$.  So in other words, if the spatial 3-metric for the regime that we live in today does not change in time, 
then we must be in a configuration of balance between the scalar field kinetic and potential energy, which would mean that they are both constant.  Using $\pi=\dot\phi$, the 
equations of motion become

\begin{equation}
{d \over {dt}}f=2\sqrt{V}\hbox{cos}\bigl[(1/2)\hbox{ln}(1/2)i(V\pi)^{-2/3}f^2\bigr]
\end{equation}

and

\begin{equation}
{{d^2} \over {dt^2}}\phi=-2if^3\Bigl({{iV\pi} \over {f^3}}\Bigr)^{2/3}\Bigl({{d\sqrt{V}} \over {d\phi}}\Bigr).
\end{equation}

\noindent
The equation for $\phi$ can be rewritten

\begin{equation}
\label{PAP13}
{{d^2} \over {dt^2}}\phi=-21f(i\pi)^{2/3}V^{2/3}{{d\sqrt{V}} \over {d\phi}}=-{6 \over 5}(i)^{5/3}\pi^{2/3}{{dV^{5/3}} \over {d\phi}}.
\end{equation}

\noindent
Multiplying both sides of (\ref{PAP13}) by $\dot\phi$ and using $\pi=\dot\phi$ we have

\begin{equation}
{d \over {dt}}({1 \over 2}\dot{\phi}^2)=-{6 \over 5}(i)^{5/3}f\dot{\phi}^{2/3}{{dV^{5/3}} \over {dt}},
\end{equation}

\noindent
or, rearranging,

\begin{equation}
\dot{\phi}^{-2/3}{d \over {dt}}(\dot{\phi}^2)=-{12 \over 5}(i)^{5/3}f{{dV^{5/3}} \over {dt}}={3 \over 2}{d \over {dt}}(\dot{\phi}^2)^{2/3}.
\end{equation}

\noindent
So the end result is

\begin{equation}
{{d(\pi^{4/3})} \over {(dV^{5/3})}}=-{8 \over 5}(i)^{5/3}f.
\end{equation}

\noindent
If $f=0$ then the scalar field kinetic energy is fixed with respect to its potential energy, and must approach a numerical constant.  Hence we would have 

\begin{equation}
\hbox{lim}_{f\rightarrow{0}}\pi^2=const\longrightarrow\dot\phi=c\longrightarrow\phi=ct+d,
\end{equation}

\noindent
where $c$ and $d$ are numerical constants of integration.  $V$ depends on $\phi$, which would mean that the potential energy would change with time as well, which is a 
contradiction.  This means that we must have $c=0$ and the potential energy is fixed- therefore the scalar field is in an equilibrium configuration.  So its constant kinetic 
energy must be zero.  The end result is that in this regime the scalar field is in equilibrium.  But if the spatial 3-metric is not singluar in this zero curvature regime 
($h^{ij}=r^{-1/3}\delta^{ij}$), then it must be the case that the potential $V$ has also decayed to zero.  The Klein-Gordon field has dissipated away due to the expansion of the 
universe into its final state, labelled by the slow-roll paramter $r$.

\section{Conclusions and directions of future research}

We have illustrated a new method to construct the generalized Kodama state for the case of a Klein--Gordon scalar field coupled to quantum gravity from solution of the quantum constraints.  We have shown this for the homogeneous and isotropic connection, illustrating the salient feature of the noncommutativity of functional with time integration.  We have expressed the quantum state in terms of its arguments as defined on the final spatial hypersurface without regard to evolution within the interior of the spacetime manifold $M$, as argued applies to the full theory in \cite{EYO}.\par
\indent
An interesting relation between minisuperpsace and the full theory, besides the holographic effect that fixes the state to the final spatial boundary $\Sigma_T$, is the contribution of the matter sector of the wavefunction to the gravitational sector via the mixed partials consistency condition on the quantized theory.  The contribution is the same in both cases, indicative of a gravity-matter interaction dictated by the SQC for the $\Psi_{GKod}$.  Another relation is the existence of a basis of states labeled by an arbitrary function of the matter fields $f$, separable in minisuperspace but nonseparable in the full theory.\par 
\indent
In the case of the minisuperspace theory, we have performed some rudimentary analysis on some effects of nonperturbative quantum gravity on the semiclassical limit.  Clearly, there are predictable effects which distinguish the coupled from the free (Minkowskian) theory, based entirely on the nondegeneracy of the Ashtekar magnetic field $B^i_a$.  It would be interesting to examine, in future work, how these predictions scale relative to the (negligible) predictions of perturbative quantum gravity in metric variables below the Planck scale.  This would require a more comprehensive analysis than performed here, but would hopefully enable one to rule out various forms of the self-interaction potential $V(\phi)$ in favor of others.  For example, in the case of the quadratic potential considered, it is clear that the orthonormality of energy eigenstates is disturbed relative to the gravity-free theory, a significant effect.  If such states are indeed orthogonal, then an input from quantum gravity by our method might point to the correct potential required to enforce this orthogonality.\par
\indent
We have also shown, for the case of a homogeneous and isotropic CDJ matrix, a method to solve the constraints at the classical level $(q_0=0)$ and have given a rudimentary analysis the semiclassical orbits of the resulting spacetime.  This method, although it does not take account of quantum effects $(q_1=q_2=0)$, does provide some predictions which can be made at the semiclassical level for a WKB state.  We hope to perform similar examinations in general in the future for the full theory.\par
\indent
Some outstanding issues and possible future directions of research include normalizability of the generalized Kodama states and expectation values of operators, as well as reality conditions and the consideration of more general matter couplings, all in anisotropic minisuperspace models and in the full theory.

\section{Acknowledgements}

\noindent
I would like to than Ed Anderson for his open-mindedness, guidance, advice and support during various lenghty discussions regarding the initial drafts of my work, as well as advice in various aspects of paper writing, as well as his patience and recommendations in reading over the final version.  I would also like to thank various other members of DAMTP.

\bigskip
\section{Appendix A: Roots of the cubic polynomial in closed form}
\par
\medskip
\indent

We illustrate in detail the well-known methods of Cardano and Ferrari for finding roots of a cubic polynomial, due to its reoccurence in the majority of the models that will come 
under consideration via our new method for constructing generalized Kodama states.  Starting from 

\begin{equation}
{{\hbox{det}B} \over {\hbox{det}{\widetilde{\sigma}}}}+\Bigl({1 \over 6}V(\phi)+{{\pi}^2 \over {12\hbox{det}{\widetilde{\sigma}}}}\Bigr)^3=0
\end{equation}

\noindent
we making the definitions

\begin{equation}
{V \over 6}=\kappa;~~~{{\pi^2} \over 12}=\rho;~~~\hbox{det}\widetilde\sigma={1 \over x},
\end{equation}

\noindent
then we must solve

\begin{equation}
(\kappa+\rho{x})^3+x\hbox{det}B
=\rho^3{x}^3+3\rho^2\kappa{x}^2+3\rho\kappa^2{x}+\kappa^3+x\hbox{det}B=0
\end{equation}

\noindent
Dividing by $\rho^3$ and making the replacement $x=y+r$ we have

\begin{equation}
(y+r)^3+3\bigl({\kappa \over \rho}\bigr)(y+r)^2+3\bigl({\kappa \over \rho}\bigr)^2{(y+r)}+\bigl({\kappa \over \rho}\bigr)^3
+(y+r)\Bigl({{\hbox{det}B} \over \rho^3}\Bigr)=0.
\end{equation}

\noindent
Expanding this out we find the following equation

\begin{equation}
\label{CUBIC}
y^3+3\bigl(r+{\kappa \over \rho}\bigr)y^2+\Bigl[3\bigl(r+{\kappa \over \rho}\bigr)^2+{{\hbox{det}B} \over \rho^3}\Bigr]y
+\bigl(r+{\kappa \over \rho}\bigr)^3+r\Bigl({{\hbox{det}B} \over \rho^3}\Bigr)=0
\end{equation}

\noindent
We choose $r=-(\kappa/\rho)$ so that the $y^2$ term in (\ref{CUBIC}) drops out and we end up with the equation

\begin{equation}
y^3+\Bigl({{\hbox{det}B} \over \rho^3}\Bigr)y-{\kappa \over {\rho^4}}\hbox{det}B=0
\end{equation}

\noindent
Making the definitions
\begin{equation}
p={{\hbox{det}B} \over \rho^3},~~~q={\kappa \over {\rho^4}}\hbox{det}B
\end{equation}

\noindent
we obtain the equation

\begin{equation}
y^3+py=q
\end{equation}

\noindent
making the substitution

\begin{equation}
y=\Bigl({{4p} \over 3}\Bigr)^{1/2}m
\end{equation}

\noindent
we transform this into

\begin{eqnarray}
m^3+{3 \over 4}m=q\Bigl({{4p} \over 3}\Bigr)^{-3/2}
=\Bigl({{4\hbox{det}B} \over {3\rho^3}}\Bigr)^{-3/2}\Bigl({\kappa \over {\rho^4}}\hbox{det}B\Bigr)\nonumber\\
=\Bigl({3 \over 4}\Bigr)^{3/2}\Bigl(\hbox{det}^{-1/2}B\Bigr)\kappa\rho^{1/2}\nonumber\\
=\Bigl({3 \over 4}\Bigr)^{3/2}\kappa\sqrt{{\rho \over {\hbox{det}B}}}
\end{eqnarray}

\begin{equation}
=\Bigl({3 \over 4}\Bigr)^{3/2}{V \over {6}}\sqrt{{(\pi^2/12) \over {\hbox{det}B}}}
={1 \over 32}V(\phi)\sqrt{{{\pi^2} \over {\hbox{det}B}}}
\end{equation}

\noindent
We are being thorough with the terms so as not to lose any appropriate constants.  We end up with

\begin{equation}
m^3+{3 \over 4}m={1 \over 32}V(\phi)\sqrt{{{\pi^2} \over {\hbox{det}B}}}
\end{equation}

\noindent
Using the trigonometric identity

\begin{equation}
\hbox{sinh}^3{\theta}+{3 \over 4}\hbox{sinh}\theta={{\hbox{sinh}(3\theta)} \over 4}
\end{equation}

\noindent
we can make the identification

\begin{equation}
{{\hbox{sinh}(3\theta)} \over 4}={1 \over 32}V(\phi)\sqrt{{{\pi^2} \over {\hbox{det}B}}}
\longrightarrow{m}=\hbox{sinh}\theta
=\hbox{sinh}\Bigl[(1/3)\hbox{sinh}^{-1}\Bigl({1 \over 8}V(\phi)\sqrt{{{\pi^2} \over {\hbox{det}B}}}\Bigr)\Bigr]
\end{equation}

\noindent
Recalling

\begin{equation}
y=\Bigl({{4p} \over 3}\Bigr)^{1/2}m=y=\Bigl({{4\hbox{det}B} \over {3\rho^3}}\Bigr)^{1/2}m
=\Bigl({{4\hbox{det}B} \over {3(\pi^2/12)^3}}\Bigr)^{1/2}m
=48\Bigl({{\hbox{det}B} \over {\pi^6}}\Bigr)^{1/2}m
\end{equation}

\noindent
we have

\begin{equation}
y=48\Bigl({{\hbox{det}B} \over {\pi^6}}\Bigr)^{1/2}
\hbox{sinh}\Bigl[(1/3)\hbox{sinh}^{-1}\Bigl({1 \over 4}V\sqrt{{{\pi^2} \over {\hbox{det}B}}}\Bigr)\Bigr]
\end{equation}

\noindent
A quick review of the mass dimensions of all quantities indicates that the formula is dimensionally consistent

\begin{equation}
[V]=2,~~[\pi^2]=2,~~[\hbox{det}B]=6
\end{equation}

\noindent
since the energy momentum tensor has two mass dimensions due to multiplication by $G$ in Einstein's equations, 
and the curvature has mass dimension two (being the derivative of a 'vector' field of mass dimension one.
Recalling the original equation

\begin{equation}
\widetilde{\sigma}^k_c=-{{B^k_c} \over {\kappa+\rho{x}}}
\end{equation}

\noindent
and that

\begin{equation}
x=y-{\kappa \over \rho}\longrightarrow\kappa+\rho{x}=\rho{y}={{\pi^2} \over 12}y=\Lambda_{eff}
={1 \over {2X}}\hbox{sinh}\bigl[(1/3)\hbox{sinh}^{-1}(VX)\bigr],
\end{equation}

\noindent
where $\Lambda_{eff}$ is the 'effective' cosmological constant, which is now field-dependent due to the effects of matter, and $X$ is given by 
(here we will take into account the proper regimes arising from the cubic equation, but for the purposes of future publications we will normally refer to the first one involving
the $\hbox{sinh}$ function)

\begin{equation}
X={1 \over 8}\Bigl({{\pi^2} \over {\hbox{det}B}}\Bigr)^{1/2}
\end{equation}

\begin{eqnarray}
\Lambda_{eff}[X]={1 \over {2X}}\hbox{sinh}\bigl[(1/3)\hbox{sinh}^{-1}(V{X})\bigr], ~~{{\hbox{det}B} \over {\rho^3}}>{0};\nonumber\\
{1 \over {2X}}\hbox{cosh}\bigl[(1/3)\hbox{cosh}^{-1}(V{X})\bigr], ~~{{\hbox{det}B} \over {\rho^3}}<{0},
~~X\geq{1};\nonumber\\
-{1 \over {2X}}\hbox{cosh}\bigl[(1/3)\hbox{cosh}^{-1}(V{X})\bigr], ~~{{\hbox{det}B} \over {\rho^3}}<{0},
~~X\leq{-1};\nonumber\\
{1 \over {2X}}\hbox{cos}\bigl[(1/3)\hbox{cos}^{-1}(V{X})\bigr], ~~{{\hbox{det}B} \over {\rho^3}}<{0},
~~\Bigl\vert{X}\Bigr\vert<{1};\nonumber\\
\end{eqnarray}

These possibilities take into account, for the various combinations of $\hbox{det}B$ and $\Omega$, whether the state falls into the Lorentzian (oscillatory) or into
the Euclidean (tunelling) regions.  In any event, $\Lambda_{eff}[X]$ appears to be a slowly varying function of $X$.  Using $\hbox{sinh}(x)\sim{x}$ for small $x$ and 
$\hbox{sinh}(x)\sim{1 \over 2}e^x$ for large $x$ we can see the dependence of $\Lambda_{eff}$ in these parameter regimes.

\begin{equation}
\lim_{X\rightarrow{0}}\Lambda_{eff}(X)={1 \over 6}V~~~;~~~\lim_{X\rightarrow\infty}\Lambda_{eff}(X)=V^{1/3}(2X)^{-2/3}
\end{equation}

\end{document}